\documentclass[twocolumn,showpacs,prl,preprintnumbers]{revtex4}
\usepackage{graphicx}

\def\bea{\begin{eqnarray}}
\def\eea{\end{eqnarray}}

\begin{document}

\title{Elasticity of Stiff Biopolymers}
\author{Abhijit Ghosh, Joseph Samuel and Supurna Sinha}
\affiliation{Raman Research Institute, Bangalore, India 560080}
\date{\today}
\begin{abstract}

We present a statistical mechanical study of stiff polymers,
motivated by experiments on actin filaments and the considerable
current interest in polymer networks. We 
obtain simple, approximate analytical forms for the force-extension 
relations and
compare these with numerical treatments. We note the important
role of boundary conditions in determining force-extension
relations.
The theoretical predictions presented here can be
tested against single molecule experiments on neurofilaments 
and cytoskeletal filaments like actin and microtubules. 
Our work is motivated by  
the buckling of the cytoskeleton of a cell
under compression, 
a phenomenon of interest to biology.

\end{abstract}

\maketitle
In recent years, statistical mechanics of semiflexible polymers
has emerged as an active area of research. 
This has been triggered by single molecule experiments designed to 
understand the role of elasticity of these polymers. Elastic properties
of polymers are of  importance in biology as in the structure 
of the cytoskeleton, a biopolymer network which controls cell mechanics 
\cite{CELL,pramod}. The parameter
which determines the stiffness of a polymer is $\beta$, the ratio
of its contour length $L$ to the persistence length $L_P$. While
the entire range of $\beta$ is of biological interest, 
in this paper we focus attention on rigid filaments such as
actin filaments and microtubules which constitute the
cytoskeletal structure and serve as tracks for motor proteins
like myosin and kinesin\cite{pramod,nelsonbook}.
Recently,
filaments of intermediate rigidity like neurofilaments have also been 
studied in some detail\cite{neuro}.
It has been shown that some remarkable features
of single stiff filament bending response are relevant to crosslinked
biofilament networks \cite{pramod}. 
A good understanding of the elastic properties of 
biopolymers at the single molecule
level, is essential to a study of polymer networks.

There are two classes of experiments which probe the 
elasticity of single biopolymers.
In one class of experiments\cite{bust} 
a semiflexible polymer molecule is pulled and stretched 
to study its``equation of state'' 
by measuring its extension as a
function of applied
force. In the other class of experiments, 
one tags the ends with fluorescent 
dye\cite{ott,loic} to 
determine the distribution
of end-to-end distances. 
Such experimental studies provide valuable insight into the 
mechanical properties of semiflexible polymers.
A good theoretical model is needed to correctly interpret these 
experiments. A simple and popular model which captures much of 
the physics is the Worm Like Chain model\cite{kratky}.

 
In this paper we analyze the bending degrees of freedom of a stiff polymer
where at least one end of the polymer 
is {\it clamped}. By this it is meant that 
the tangent vector at this end is kept in a fixed direction. 
Just as in the case of a stretched polymer\cite{bend},
the tangent vector of a stiff polymer executes
small wanderings around this fixed direction. 
The theoretical analysis for the statistical mechanics 
of a stiff polymer, clamped at least at one end is similar to that of a
polymer in the high stretch limit. We refer to this approximation
as the paraxial approximation. This approximation has been previously
used to study the elasticity of twist storing flexible 
stretched polymers\cite{bouchiat,nelson,writhe,dnael} 
in the paraxial worm like chain model (PWLC model). 
In an earlier paper 
\cite{bend}
we had an ``exact'' numerical scheme for a semiflexible polymer with free
boundary conditions for the end tangent vectors. 
The main difference in the numerical 
scheme presented here is that we impose boundary conditions 
on the tangent vectors at the ends. As a general rule,  
in the stiff regime the convergence 
of the numerical scheme is poorer and therefore one needs to use 
larger matrix sizes.
The search for simple analytic forms to describe the elastic properties
of stiff polymers is therefore well motivated.

For stiff polymers, the experimentally measured mean values 
crucially depend \cite{inequ} on the precise choice 
of the 
ensemble. 
This is due to finite size fluctuation effects, which are 
entirely absent in the elasticity of a classical rod. 
For 
instance,
one gets qualitatively distinct features in force-extension curves 
depending
on whether the force or the extension is held constant in an
experimental setup \cite{dabhi,kreuz,gaut}. This is an aspect of stiff 
polymer
statistical mechanics which is both theoretically challenging and 
experimentally significant. In this paper, we remain throughout
in the Gibbs ensemble, where the applied force is held fixed
and we measure the mean extension.

The organization of this paper is as follows. We first present 
results based on an ``exact'' numerical 
scheme for stiff polymers for two different boundary conditions, one
in which both ends are clamped and the other in which one end is 
clamped and the other end is free.
We then present simple analytical forms for these two cases.
Our main results are displayed in Figs. 1-3 comparing
the numerical scheme with the analytic formulae for force-extension
relations. 
Finally, we end the paper with a concluding 
discussion of the buckling of stiff polymers and the consequent breakdown 
of the paraxial approximation.
                                                                               
Our starting point is the
{\it Worm Like Chain (WLC) model} in which
the configuration ${\cal C}$ of the polymer is
described by a  space curve ${\vec x}(s)$, with $s$ the arc-length
parameter ($0\le s \le L$) ranging from $0$ to $L$, the contour
length of the
polymer.
The tangent vector ${\hat t}=d{\vec x}/ds$  to the curve is a unit vector
\begin{equation}
{\hat t}.{\hat t}=1
\label{unit}
\end{equation}
and the curvature of the polymer is given by $\kappa=|d{\hat t}/ds|$.

One can study the case of stiff polymers using a combination of
analytical and numerical techniques \cite{bend}. Let one end of
the polymer be fixed at the origin and a stretching force $F$ be applied
in the ${\hat z}$ direction, which we refer to as the 
north pole of the sphere of directions. 
Introducing
a dimensionless force variable $f=\frac{F L_p}{k_B T}$, 
where $k_B T$ is the thermal energy
we can express the partition 
function $Z(f)$ as 
\begin{equation}
Z(f)={\cal N}\int{\cal D}[{\hat t}(\tau)]
e^{-\int_0^\beta d\tau[1/2(d{\hat t}/d\tau)^2
-f{\hat t}_z]}
\label{path2}
\end{equation}
where $\beta =L/L_p$.
Eq. (\ref{path2}) can be interpreted as the path integral
representation
for the kernel of a {\it quantum} particle on the
surface of a sphere at inverse temperature $\beta$.
Thus we can express $Z(f)$ as the quantum amplitude to go from
an initial tangent vector $\hat t_A$ to a final tangent vector $\hat t_B$
in imaginary time $\beta$ in the presence of an
external potential $-f \cos\theta$:
\begin{equation}
Z(f,\hat t_A,\hat t_B)=<\hat t_A| \exp[-\beta {\hat H}_f] |\hat t_B> 
\label{operz}
\end{equation}
The Hamiltonian ${\hat H}_f = -\frac{\nabla^2}{2} -f \cos\theta $ is
that of a rigid rotor\cite{Marko} in a potential. In the absence of
a force, the 
free Hamiltonian is  $H_0 = -\frac{1}{2} \nabla^2$.
By choosing a standard basis in which
$H_0$ is diagonal we find that
$H$ is a symmetric tridiagonal matrix with diagonal
elements
$H_{l, l}={l(l+1)}/{2}$
and superdiagonal  elements
$H_{l, l+1}=f (l+1)\sqrt{1/((2l+1)(2l+3))}$. 
Inserting a complete set of eigenstates of the free Hamiltonian
into Eq. (\ref{operz}), we find $Z(f,\hat t_A,\hat t_B)=$
\begin{equation}
\sum_{m,n}<\hat t_A|\psi_n> <\psi_n|\exp[-\beta 
{\hat H}_f] |\psi_m><\psi_m|\hat t_B> 
\label{operdiag}
\end{equation}
where ${\hat M}^{f}=\exp[{-\beta {\hat H}_f}]$. From this general form, we 
can compute
the partition function in the present cases of interest.

In Ref. \cite{bend} we had studied the elastic properties of polymers with
free boundary conditions: the directions of the tangent vectors at both
ends were integrated over. In the present paper, we will fix the tangent
vector at the ends (one or both) to lie along the $\hat{z}$ direction. To 
implement this numerically, we have to evaluate the eigenfunctions 
in Eq. (\ref{operdiag}) at this value of ${\hat{t}}$.

{\it $(i)$both ends clamped}: $\hat t_A=\hat t_B=\hat z$. While a complete
set of eigenstates are labelled by $(l,m)$, only the $m=0$ terms 
contribute here because of azimuthal symmetry and we have
\begin{equation}
Z(f,\hat z,\hat z)=\sum_{l,l'}U_l M^{f}_{l,l'} U_{l'}=U.M.U
\label{bothends}
\end{equation}
where $U_l=\sqrt{\frac{2l+1}{4\pi}}$.

{\it $(ii)$one end clamped}: Integrating Eq. (\ref{operdiag}) over $\hat t_B$,
we find that 
\begin{equation}
Z(f,\hat t_A)=\sum_{l}U_l M^{f}_{l0}=(U.M^{f})_0
\label{oneend}
\end{equation}
Both Eqs. (\ref{bothends}) and (\ref{oneend}) are suitable for numerical 
implementation. $H_f$ is an infinite symmetric matrix. We truncate it to
finite order $N$ and choose $N$ large enough to attain the desired 
accuracy \cite{math}.

While this numerical method is effective, it has a limitation in 
describing stiff polymers due to the poor convergence of statistical
sums in Eq. (\ref{operdiag}). For stiff polymers a convenient and accurate 
analytical approximation scheme can be developed as shown below.

For a stiff polymer with one end clamped along the 
${\hat z}$ direction, we can approximate the sphere
of directions by a tangent plane
at the north pole of the sphere as the
angular coordinate $\theta$ always remains small.
Introducing Cartesian coordinates $\xi_{1}=\theta \cos{\phi}$ and
$\xi_{2}=\theta \sin{\phi}$
on the tangent plane $R^2$ at the north pole
one can express the small $\theta$ Hamiltonian $H$ as $H=H_P-f$ 
where $H_P$ is
\begin{equation}
H_P=\frac{1}{2} {p_{\xi_{1}}}^2
+\frac{1}{2} {p_{\xi_{2}}}^2+\frac{f}{2}({\xi_{1}}^2+{\xi_{2}}^2);
\label{hampwlcp}
\end{equation}
Notice that $H_P$ is the Hamiltonian
of a two-dimensional harmonic oscillator with a
frequency $\omega = \sqrt{f}.$ For a single oscillator in 
real time the propagator is given by\cite{feynhibb}:
$K(\xi_i,\xi_f,T)=$ 
\begin{equation}
F(T)\exp{\frac{i\omega}{2\sin{\omega 
T}}[(\xi_i^2+\xi_f^2)\cos{\omega 
T}-2\xi_i \xi_f]}
\label{oscillator}
\end{equation} 
where $F(T)=\sqrt{\frac{\omega}{2\pi i \sin(\omega 
T)}}$.

For the sake of convenience we set $L_P = 1$ so that $\beta = L$.

{\it Case $(i)$: both end tangent vectors clamped along 
the ${\hat z}$-direction: }
Setting $\xi_i = \xi_f = 0$ in Eq. (\ref{oscillator}) and 
continuing the expression to imaginary time results in the
trigonometric functions being replaced by hyperbolic ones.
We can express the partition function $Z(f)$ as $\exp{(\beta f)}$ times 
the product of the propagators of
two independent harmonic oscillators:
\begin{equation}
Z(f) = \sqrt{f}\exp (\beta  f)/(2 \pi \sinh\big(\beta \sqrt{f}) \big).
\label{zoff1}
\end{equation}
in Euclidean time $\beta$; the free energy is 
$G(f) = -\log{Z(f)}/\beta$
\begin{equation}
=[\frac{-1}{2\beta}logf -f +\frac{1}{\beta}log(2\pi) 
+\frac{1}{\beta}log[\sinh(\beta\sqrt{f})]
\label{free}
\end{equation}
The mean extension $<\zeta>=<z>/L=-\partial G(f)/\partial f$ is given by 
(See Fig. $1$).

\begin{equation}
<\zeta> = 1 +  1/(2 \beta f) - \coth(\beta \sqrt{f})/(2 \sqrt{f}).
\label{fer}
\end{equation}
where $<\zeta>$ is the ${\hat z}$ component of the extension
(or the end-to-end distance vector).
Note in Fig. 1 that the analytical form agrees with the numerical scheme 
to an accuracy of about $1\%$.
                                                        
{\it Case $(ii)$: A stiff polymer with one end tangent vector pointing 
along 
the ${\hat z}$-direction and the other end free: }

In this case the propagator
for the harmonic oscillator has to be integrated over
the final coordinates $\xi_f$
and evaluated at $\xi_i = 0$. 
The partition function in
this case turns out to be
\begin{equation}
Z(f) =   \exp (\beta  f)/\cosh(\beta \sqrt{f}).
\label{zoff}
\end{equation}
From the expression of the partition function,
we get the free energy 
\begin{equation}
G(f) =  -f + \frac{1}{\beta}log[\cosh(\beta{\sqrt{f}})]
\label{goff}
\end{equation}

and differentiate
it with respect to $f$ to get the
force - extension relation -
\begin{equation}
<\zeta> = 1 - \tanh(\beta \sqrt{f})/(2 \sqrt{f}).
\label{zxtn}
\end{equation}

Note that even at zero force, there is a nonzero extension, because of
the boundary condition and the stiffness of the polymer. 
Fig. $2$ shows a comparison of the force extension curves for 
the two boundary conditions. We find, as expected, that for 
the same force, the extension is larger for the more 
constrained boundary condition [Case(i)] compared to a less
constrained one [Case(ii)]. 
For positive forces the paraxial approximation is very good and 
the forms are displayed in Eqs. (\ref{fer}) and (\ref{zxtn}).
For large positive forces they become 
$<\zeta> = 1 - \frac{1}{2 \sqrt{f}}$ \cite{Marko,bend} as 
expected. 

For negative forces, the hyperbolic functions appearing in 
Eqs. (\ref{zoff1}) and (\ref{zoff}) go over to circular functions.
For instance, for the case in which 
both end tangent vectors are clamped along the ${\hat z}$
direction, for negative $f$, 
our simple analytical 
form for the partition function reads
\begin{equation}
Z(f)=\frac{\sqrt{-f}e^{\beta f}}{2\pi \sin(\beta\sqrt{-f})}
\label{zofg}
\end{equation}
and varies continuously with $f$ as $f$ ranges from positive to 
negative values. 
As the compressive force is increased, we find that at a critical 
value of the force $f$ the extension $<\zeta>$ spontaneously 
decreases. This is the analogue here of the classical Euler
buckling instability which occurs for rods.  

Consider the mean extension versus force relation [Eq.(\ref{fer})]
for negative forces (i.e. for compressive forces). 
For negative values of forces Eq.(\ref{fer}) reduces to 
\begin{equation}
<\zeta> = 1 +  1/(2 \beta f) - \cot(\beta \sqrt{-f})/(2 \sqrt{-f}).
\label{ferneg}
\end{equation}
which can be rewritten in the form
\begin{equation}
<\zeta> = 1 +  \beta u(x).
\label{uofx}
\end{equation}

where $x=\beta\sqrt{-f}$ and 
$$u(x)=\frac{\cot(x)}{2x} -\frac{1}{2x^2}$$

The criterion for the onset of the buckling instability is the divergence 
of $\partial <\zeta>/\partial f$. From Eq.(\ref{uofx}) this is 
equivalent to the divergence of $\partial u/\partial x$, which 
takes place at a value of $x_c = \pi$.
This gives us the following expression for the critical 
force for buckling\cite{landau}:
$$f_c ={-\Big(\frac{\pi}{\beta}\Big)^2}$$  

Because of the quadratic dependence, the 
compressive force needed to buckle a polymer rises sharply with 
stiffness.   
The mean extension versus force curves displayed in Fig. $3$
demonstrate the phenomenon of buckling. As expected, we notice 
that as $\beta$ goes up, the magnitude of the critical force $f_c$ needed 
to buckle 
the polymer goes down.   

A stiff polymer is energy dominated and its buckling is
very similar to that of a classical rod subject to identical
boundary conditions and a compressive force. The effect of 
thermal fluctuations is to slightly ``round off'' the transition
from the straight to the buckled configuration. This is due to 
thermally activated processes that permit the polymer to overcome
the elastic energy barrier. As a result the buckling force for
a stiff polymer is slightly smaller in magnitude than the 
$f_c$ given above.


There is a long history of the use of path integrals in the study of polymers
\cite{Schulman,Edwards}. 
Such methods have 
been used in the study of elasticity of semiflexible 
polymers\cite{bend,Kleinert,saito,yam}. 
This connection between path 
integrals in quantum mechanics and statistical mechanics of 
polymers enables us to import ideas back and forth between 
these two distinct domains. 
The main point of this paper 
is that standard results in  path integrals give us
new results for stiff biopolymers. Our main results are 
contained in the analytic forms 
displayed in Eqs. (\ref{zoff1} -\ref{zxtn}) and Figs. (1-3).

In this paper we have theoretically studied the elasticity of stiff 
biopolymers. We have studied some cases with boundary 
conditions realizable in single molecule experiments.
By attaching a magnetic bead to an end of the polymer, one can 
apply forces using magnetic field gradients and torques using
magnetic fields. By such techniques one can impose a variety of 
boundary conditions on the polymer including the ones discussed 
here. 
Recent studies have shown \cite{pramod} that the elastic behavior of such 
a biopolymer at the single molecular level 
affects the 
elastic properties of a biopolymer network.
This is much like the way 
the structural stability of a roof is determined by
the rigidity of its rafters.
In a cytoskeletal structure the end tangent 
vectors of the stiff biopolymers that make up the structure are pinned.
A cytoskeleton can be viewed as a replica of a large number $N$
of semiflexible polymers. By studying the elastic properties of a single
polymer constituting such a network, we can draw conclusions regarding 
the stability of the $N$ polymer cytoskeletal structure. 
Here we have presented 
closed form simple analytical expressions for 
force-extension relations for a single stiff filament which can be 
tested against single molecule experiments. 
These analytical results are new and are expected to shed light on 
the structural stability of the $N$ polymer cytoskeletal structure.  

 We have also considered the case in which one end of a stiff polymer is clamped  and the other end is free. This is a boundary condition 
that is more natural to an experimental setup for measuring the 
end-to-end distance distribution $P(\zeta)$ of a polymer via imaging of 
a polymer tagged with fluorescent dye. In fact one can construct a 
force-extension curve from the 
experimental data of $P(\zeta)$ versus $\zeta$. 
We have theoretically analyzed this case and made 
predictions for experiments in this case as well. 
As in the earlier case, in this case also 
we have a simple analytic form. 
This is another new result.

In future, we would like to investigate 
buckling of stiff filaments 
like actin in greater detail . 
This is an issue that is of relevance at the 
single molecular level as well as at the level of a biopolymer network 
like the cytoskeletal structure and is expected to shed light on its 
structural stability\cite{pramod,rbc} and collapse under stress.
The stiffness and collapse of the cytoskeletal structure of a red blood
cell\cite{rbc} has a direct connection to its functional 
aspects and used for instance, as a diagnostic for detection of sickle 
cell anaemia.  
In studying the 
cytoskeletal structure it would be most useful to have 
a good understanding of the individual polymers that make up the
structure. 
Simple analytic forms give valuable insight into a problem and we expect
the analytic results 
presented here to provide some fresh impetus to this rapidly growing
field of semiflexible polymers.  

\begin{figure}[p]
\includegraphics[height=4.0cm,width=8.2cm]{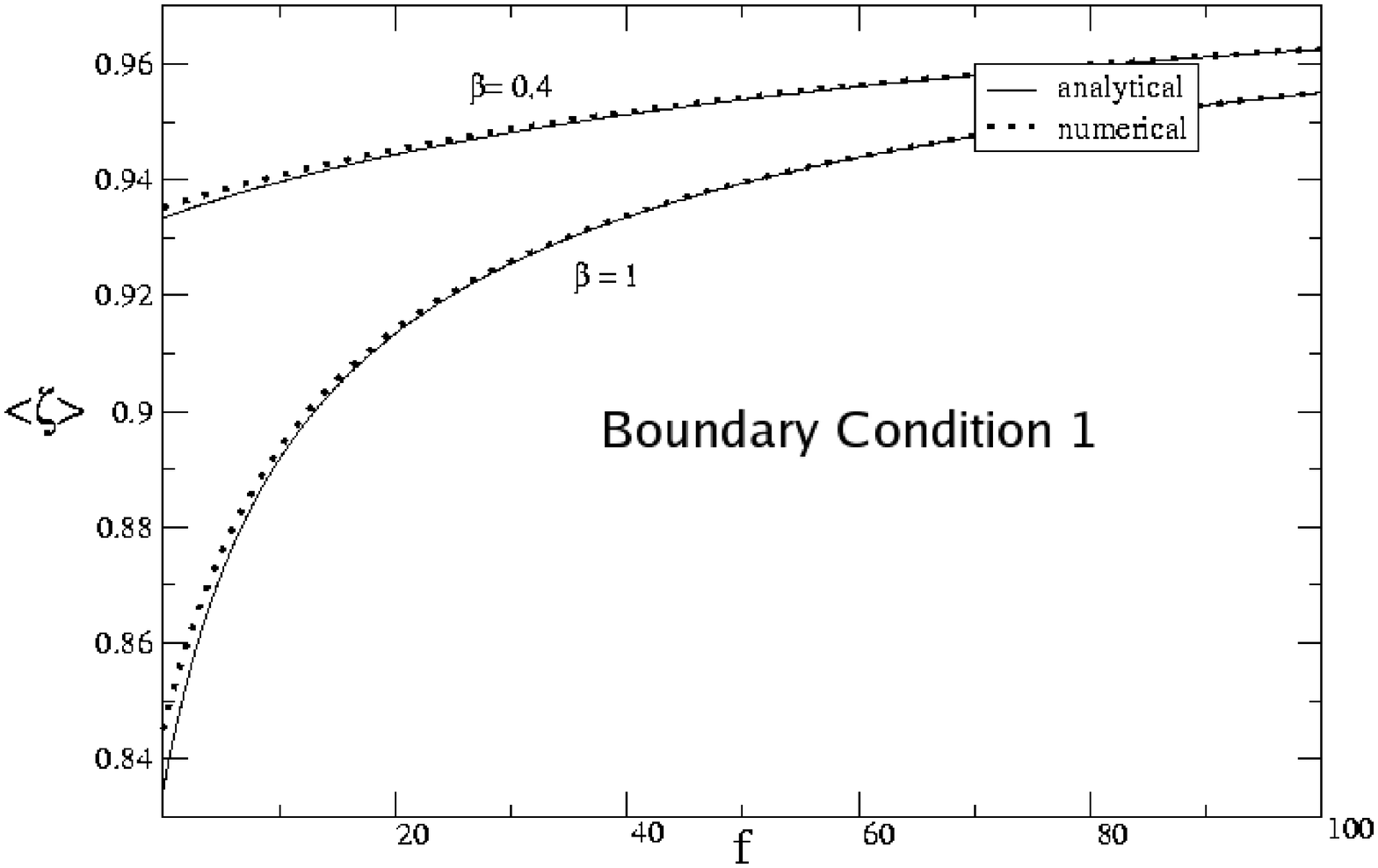}
\caption{The mean extension is plotted against the force $f$ 
 for $\beta = .4, 1$ for a setup with both ends clamped.}
\label{Fig. $1$}
\end{figure}

\begin{figure}[p]
\includegraphics[height=5.0cm,width=8.2cm]{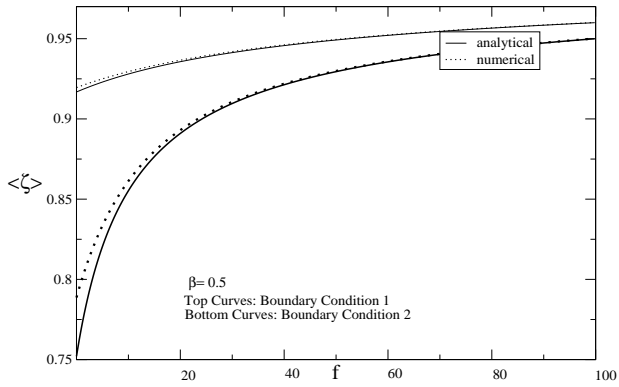}
\caption{The figure demonstrates that force-extension relations depend on 
the boundary conditions. Note that as expected, for a given force, the 
extension is greater for the case where {\it both} ends are 
clamped in the ${\hat z}$ direction.}
\label{Fig.$2$}
\end{figure}
\begin{figure}[p]
\includegraphics[height=6.0cm,width=9.0cm]{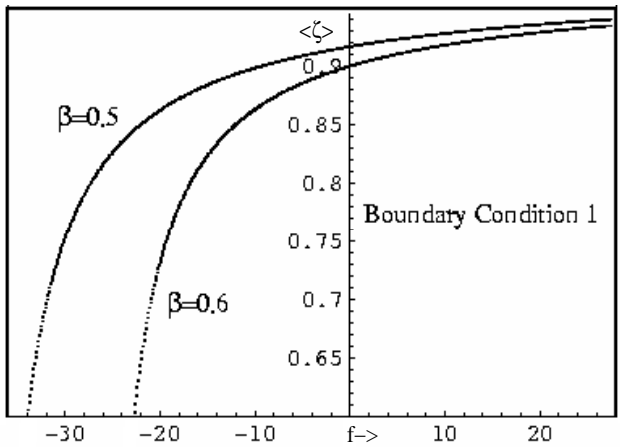}
\caption{ Figure shows buckling {\it i.e.} spontaneous
decrease in extension under a compressive force for a stiff
polymer with end tangent vectors clamped
for $\beta = .5$ and $\beta = .6$.
Note that buckling takes place at a smaller magnitude of the 
compressive force $f$ for a larger $\beta$. 
}
\label{Fig. $3$}
\end{figure}


\normalsize
                                                                                
                                                                                
\begin{thebibliography}{**}
\bibitem{CELL} B. Alberts et al, {\it Molecular Biology Of The Cell}
(Garland Publishing, New York, 1994), 3rd ed.
\bibitem{pramod}
P. Fernandez et al, Biophysical Journal {\bf 90}, 1 (2006) and 
references therein.
\bibitem{nelsonbook} P. Nelson, 
Biological Physics: Energy, Information, Life, (W. H. Freeman, 2003).
\bibitem{neuro} 
Z. Dogic et al, {\it Phys. Rev. Lett.} {\bf 92}, 125503 (2004).
\bibitem{bust} C. Bustamante et al,
Current Opinion in Structural Biology {\bf 10}, 279 (2000).
\bibitem{ott}
A. Ott, M. Magnasco, A. Simon and A. Libchaber,
{\it Phys. Rev.} {\bf E48}, 
R1642 (1993).
\bibitem{loic}
L. Le Goff, O. Hallatschek, E. Frey and F. Amblard, 
{\it Phys. Rev. Lett.} {\bf 89}, 258101 (2002).
\bibitem{kratky}
O. Kratky and G. Porod, {\it Rec. Trav. Chim. Pays-Bas.} {\bf 68} 
1106-1123 (1949).
\bibitem{bend} J. Samuel and S. Sinha,
{\it Physical Review E}, {\bf 66} 050801(R) (2002);
S. Stepanow and G. M. Sch\"utz,
{\it Europhysics Letters}, {\bf 60} 546 (2002).
\bibitem{bouchiat}
C. Bouchiat  and M. Mezard, {\it Phys. Rev. Lett.} {\bf 80}, 1556 
(1998).  
\bibitem{nelson} 
Moroz J D and Nelson P, {\it Proc. Natl. Acad. Sci. USA} {\bf 94}, 14418 
(1997).
\bibitem{writhe} S. Sinha,
{\it Physical Review E} {\bf 70} 011801 (2004).
\bibitem{dnael} J. Samuel, S. Sinha and A. Ghosh
{\it Journal of Physics: Condensed Matter}, {\bf 18} S253 (2006).
\bibitem{inequ}
S. Sinha and J. Samuel, {\it Physical Review E} {\bf 71}, 021104 (2005).
\bibitem{dabhi} A. Dhar and D. Chaudhuri, {\it Phys. Rev. Lett.} {\bf 89}, 
065502 (2002).
\bibitem{kreuz}
H. J. Kreuzer and S. H. Payne, {\it Phys. Rev.} {\bf E 63},021906 (2001).
\bibitem{gaut} P. Ranjith, P.B. Sunil Kumar and G.I. Menon,
{\it Physical Review Letters}, {\bf 94} 138102 (2005).
\bibitem{Marko}
J. Marko and E. D. Siggia, Macromolecules {\bf 28}, 8759 (1995).
\bibitem{math}
Stephen Wolfram, The Mathematica Book, Third Edition (Wolfram Media/
Cambridge University Press,1996).
\bibitem{feynhibb}
R. P. Feynman and A. R. Hibbs, 
Quantum Mechanics and Path Integrals, (McGraw-Hill Companies, 1965).
\bibitem{landau}
L. D. Landau and E. M. Lifshitz, 
Theory of Elasticity, pg 98, Problem $2$ (Pergamon Press Ltd. (1970) 
Britain.
\bibitem{Schulman}
L. S. Schulman, Techniques and Applications of Path Integration, (Wiley interscience, 1981).
\bibitem{Edwards}
S. F. Edwards and M. Doi, The Theory of Polymer Dynamics, (Oxford University Press, Oxford, 1986).
\bibitem{Kleinert}
H. Kleinert, Path Integrals in Quantum Mechanics, Statistics, Polymer Physics and Financial Markets, (World scientific, Singapore, 2006, 4. ed.) 
\bibitem{saito}
N. Saito, K. Takahashi and Y. Yunoki {\it J. Phys. Soc. Jpn} {\bf 22}, 
219 (1967).  
\bibitem{yam}
H. Yamakawa, Helical Wormlike Chains in Polymer Solutions, 
(Springer, New York, 1997).
\bibitem{rbc}
A. Ghosh et al,
{\it Phys. Biol.} {\bf 3} 67-73 (2006).

\end{thebibliography}
\end{document}